# Automated Description Generation of Cytologic Findings for Lung Cytological Images Using a Pretrained Vision Model and Dual Text Decoders: Preliminary Study

Atsushi Teramoto[1*], Ayano Michiba[2], Yuka Kiriyama[2,3], Tetsuya Tsukamoto[4], Kazuyoshi Imaizumi[2], and Hiroshi Fujita[5]

1. Faculty of Information Engineering, Meijo University
2. School of Medicine, Fujita Health University
3. Narita Memorial Hospital
4. Oncology Innovation Center, Fujita Health University
5. Faculty of Engineering, Gifu University

*Corresponding author

## Abstract:

Objective: Cytology plays a crucial role in lung cancer diagnosis. Pulmonary cytology involves cell morphological characterization in the specimen and reporting the corresponding findings, which are extremely burdensome tasks. In this study, we propose a technique to generate cytologic findings from for cytologic images to assist in the reporting of pulmonary cytology. Methods: For this study, 801 patch images were retrieved using cytology specimens collected from 206 patients; the findings were assigned to each image as a dataset for generating cytologic findings. The proposed method consists of a vision model and dual text decoders. In the former, a convolutional neural network (CNN) is used to classify a given image as benign or malignant, and the features related to the image are extracted from the intermediate layer. Independent text decoders for benign and malignant cells are prepared for text generation, and the text decoder switches according to the CNN classification results. The text decoder is configured using a Transformer that uses the features obtained from the CNN for generating findings. Results: The sensitivity and specificity were 100% and 96.4%, respectively, for automated benign and malignant case classification, and the saliency map indicated characteristic benign and malignant areas. The grammar and style of the generated texts were confirmed correct, achieving a BLEU-4 score of 0.828, reflecting high degree of agreement with the gold standard, outperforming existing LLM-based image-captioning methods and single-text-decoder ablation model. Conclusion: Experimental results indicate that the proposed method is useful for pulmonary cytology classification and generation of cytologic findings.

# I. INTRODUCTION

Cancer statistics show that lung cancer has the highest incidence and death rate among all cancers [1]. In addition to surgery, effective methods for treating lung cancer, including radiotherapy and chemotherapy, have been developed. Early and accurate diagnosis is necessary to obtain good therapeutic outcomes with these treatment modalities. Pathological diagnosis, which involves cytological and histological analyses, plays an important role in detailed lung cancer diagnosis. In cytological diagnosis, a specimen is prepared from cells collected via bronchoscopy or other procedures, and cytotechnologists or cytopathologists uses a microscope to observe the cell nucleus, cytoplasm, and cell arrangement to distinguish benign from malignant cases and determine the histological type; the results are then described in a report. Advancements in lung cancer diagnostic techniques and the advent of the genomic medicine era have significantly transformed the role of lung cytopathology. An international reporting format for respiratory cytology in lung cytopathology has been developed through the collaboration of multiple organizations. In 2022, the World Health Organization (WHO) released the WHO Reporting System for Lung Cytopathology, which introduced a five-tiered categorization system for structuring cytopathology reports [2]. However, cytological diagnosis is time-consuming and burdensome because it is extremely complicated and requires the observation of a large number of cells under a microscope while writing reports. This study aims to develop a technology to support cell classification and diagnostic report creation in cytology as its ultimate goal. As a preliminary study toward this objective, it explores the automated generation of cytological findings from images of benign and malignant lung cells.

Various classification methods have been developed for cytology [3-9]. Regarding lung cytology, our initial study employed a simple five-layer convolutional neural network (CNN) to classify lung cancer tissue types in Papanicolaou-stained lung specimens, achieving a classification accuracy of 71.1% [4]. Subsequently, fine-tuning multiple pre-trained CNNs improved the classification accuracy to 78.9% [5]. Furthermore, pre-trained CNNs were applied to classify benign and malignant lung cells, yielding a classification accuracy of 79.2% [6]. To further enhance accuracy, a method utilizing images generated by generative adversarial networks (GANs) was proposed, which improved the classification accuracy to 85.3% [7]. Recently, studies have also explored the classification of cytology images using whole-slide images (WSIs). Gonzalez et al. reported a method for distinguishing large cell neuroendocrine carcinoma from small cell lung carcinoma using WSIs [8]. Similarly, Park et al. conducted automated detection of metastatic breast cancer in pleural effusions using WSIs, achieving high metrics with a classification accuracy of 81.1%, a sensitivity of 95.0%, and a specificity of 98.6%, which surpassed the performance of pathologists [9].

Several studies have focused on generating findings from medical images. In the study of radiographic findings generation, Wang et al. [10] combined CNN and recurrent neural network (RNN) models to generate text findings from the ChestXRay-14 dataset. Hou et al. [11] employed a CNN-Transformer model for the same task. In the field of pathology, Zhou et al. [12] used a graph neural network and Transformer to analyze histological images of urothelial papillary carcinoma, achieving high concordance with reference findings, alongside a classification accuracy of 79.0%. The above image captioning methods extract image features and generate text using a single model and data pathway, without differentiation based on whether the case is benign or malignant. However, in clinical cytology, benign and malignant cases are differentiated based on specific cellular morphology and arrangement, which are critical for accurate image findings. Each case type requires tailored descriptions, with benign cases typically requiring simpler descriptions. Existing methods fail to account for these categorical differences, potentially leading to inaccurate descriptions. To address this issue, we propose a novel approach that combines a high-performance image classification model with a text generation algorithm to produce image findings better suited for cytological diagnosis. To the best of our knowledge, this is the first method specifically designed for cytology.

In this paper, we propose a novel scheme for the diagnosis and generation of cytologic findings for lung cytological images by combining a pretrained image classifier with dual text decoders. The WHO reporting system mentioned above indicates that reports should be generated in five categories. However, as a preliminary study, we focus on generating image findings at two levels: benign and malignant.

## II. METHODS

### A. Outline

An overview of the proposed scheme is shown in Fig. 1. The vision model uses a CNN, fine-tuned with cytological images, to distinguish malignant from benign cells and extracts features for text decoders. Based on the CNN classification results, text decoders (Text Decoders #1 and #2) are invoked for benign and malignant cells, and a description of the cytologic finding is generated based on the CNN features. The saliency map generated by the Grad-CAM is also exported. To confirm the effectiveness of the proposed method, we compared it with a model with a single-text decoder and existing state-of-the-art (SOTA) image-captioning models. Cytological images are provided to the CNN, which classifies them as benign or malignant. Based on the results, the two text decoders are switched and a description is generated based on the CNN-provided image features.

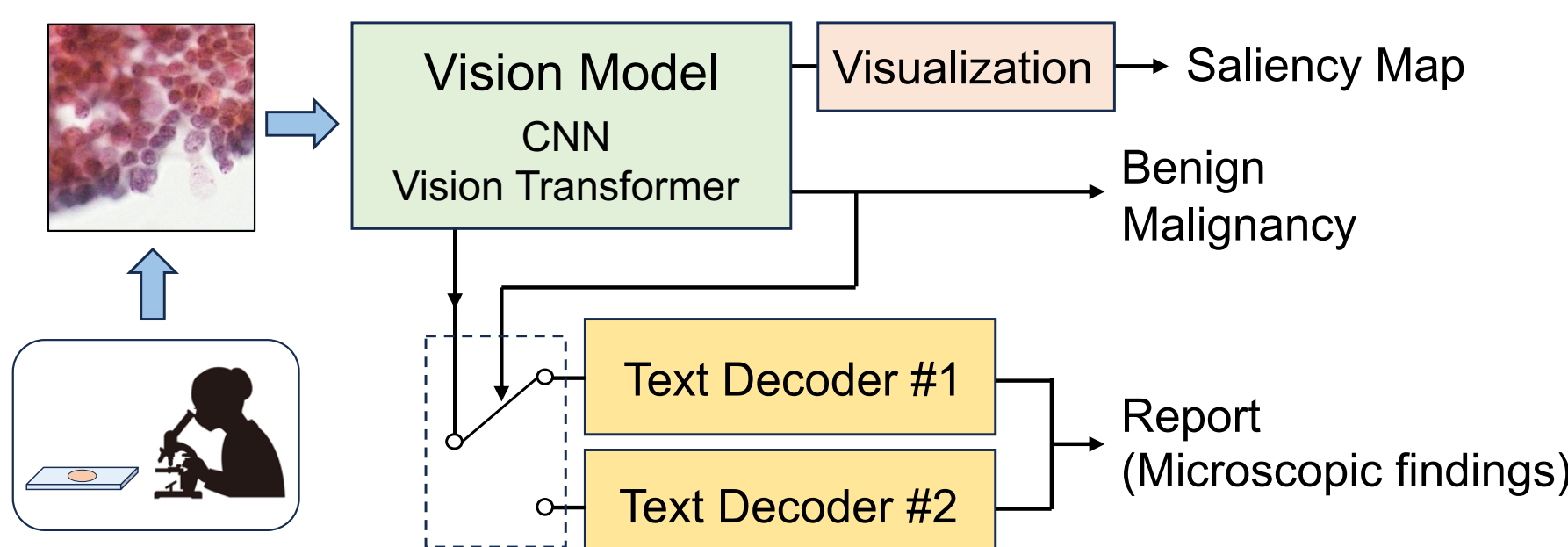

**Fig. 1** Outline of the proposed description-generation of cytologic findings.

### B. Dataset

In this study, lung cells were collected from 206 patients via interventional cytology, using either bronchoscopy or computed-tomography-guided fine-needle aspiration cytology; the collected cells consisted of 71 benign and 135 malignant cases. The malignant cases included 83 adenocarcinomas and 52 squamous cell carcinomas. These diagnoses were combined with the histological analysis of the biopsy specimens for the final determination. Biopsy tissues were collected at the same time as the cytology specimens, fixed in 10% formalin, dehydrated, and embedded in paraffin. The cytological specimens were prepared via liquid-based cytology using the BD SurePath™ liquid-based Pap test (Beckton Dickinson, Franklin Lakes, NJ, USA) and stained using the Papanicolaou method. A microscope (BX53, Olympus Corporation, Tokyo, Japan) attached to a digital camera (DP20, Olympus Corporation) was used to acquire 219 microscopic benign cell images and 460 malignant cell images in JPEG format with a size of 1280 × 960 pixels.

To construct the image dataset for description generation of cytologic findings, a cytotechnologist and cytopathologist extracted a 296 × 296-pixel patch image from areas containing cells in the original microscopic image, as shown in Fig. 2. The final dataset consisted of 801 patch images, with 325 benign and 476 malignant cell images. Microscopic findings were then prepared for these images to describe the cell type, shape of the cell nucleus, cell arrangement, and background, which represented conditions other than the target cells. The findings were described by one cytotechnologist and one cytopathologist in accordance with the World Health Organization (WHO) pulmonary cytopathology reporting system [2]. A sample of the dataset is shown in Fig. 3.

Finally, the dataset was randomly divided into training and evaluation datasets. The training dataset consisted of 270 images from 57 benign cases and 381 images from 108 malignant cases, while the evaluation dataset consisted of 55 images from 14 benign cases and 95 images from 27 malignant cases. Images of the same patient were not mixed between the training and evaluation datasets.

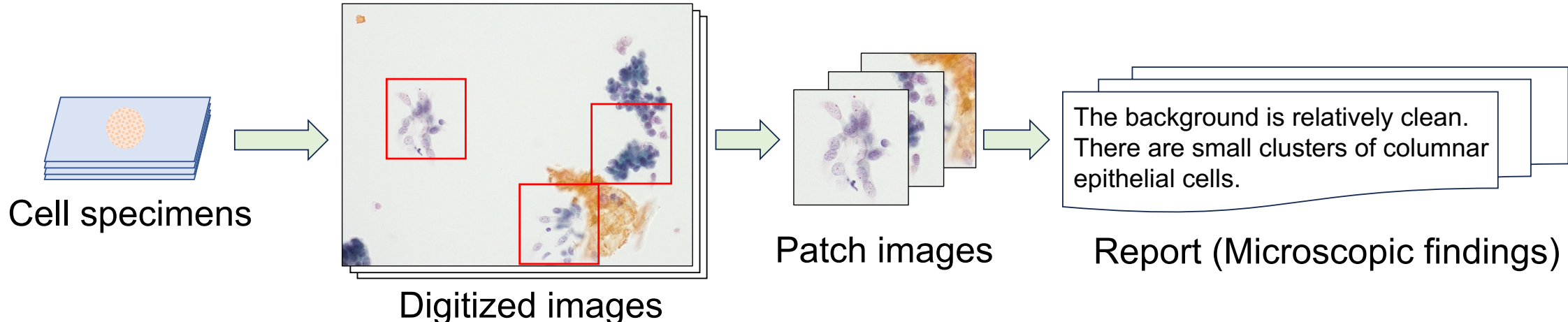

**Fig. 2** Preparation of patch images and the cytologic findings. Cytology specimens were taken under a microscope, and experts selected the areas of the specimens that should be output as a description and converted into a patch image. A description corresponding to each image was then prepared.

| | |
|---|---|
| | The background is relatively clean.<br>There is the clusters of columnar epithelial cells |
| | There are columnar epithelial cells in the inflammatory background. |
| | There are adenocarcinoma cells with hyperchromatic nucleus, nucleus with irregular shape, pale cytoplasm, promient nucleoli, filled with foamy mucus. |
| | There are squamous cell carcinoma cells with keratinized squamous cell, hyperchromatic nucleus, nucleus with irregular shape. |

**Fig. 3** Four examples of patch images and the corresponding description of cytologic findings in our dataset.

### C. Vision Model

Vision models have multiple roles. First, a given patch image is classified as benign or malignant. Fig. 4 shows the structure of the CNN-based visual model used in this study. For the CNN architecture, we introduced VGG16 [13], InceptionV3 [14], ResNet50 [15], and DenseNet121 [16], which are existing SOTA models. The parameters of these CNNs were pre-trained using the ImageNet dataset. The fully connected layers of the original CNNs were replaced with ones comprising 1024 units and 3units; and the entire network was fine-tuned using actual patch images. To augment the training data, images were rotated by 90°.

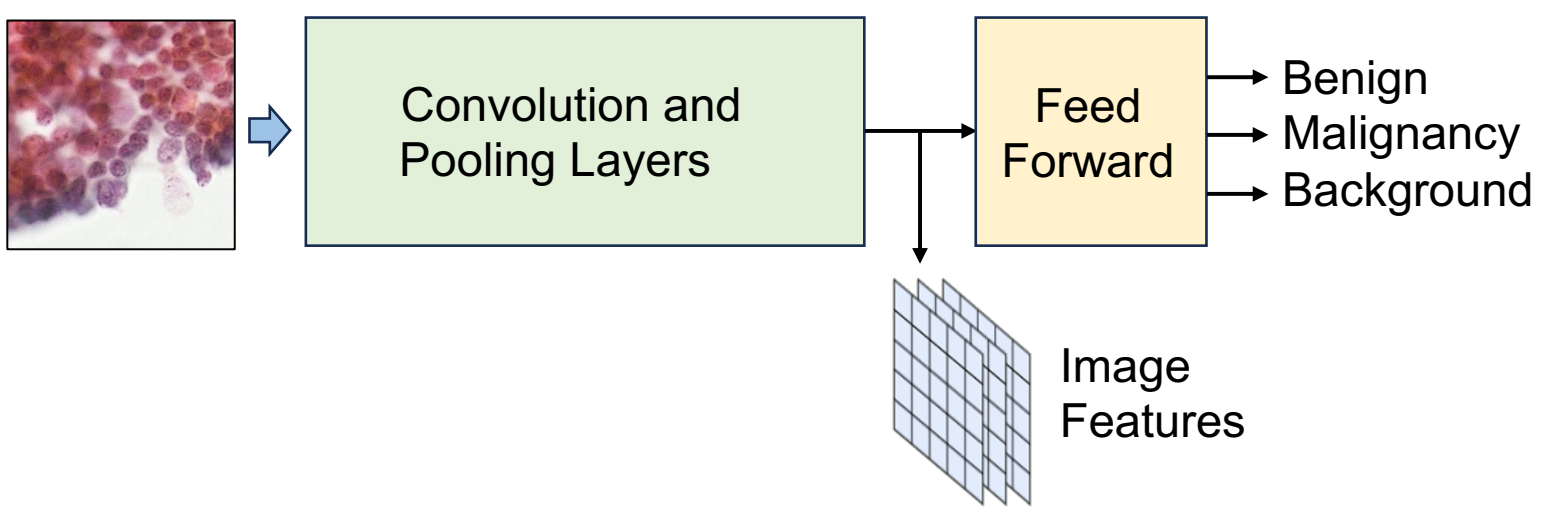

**Fig. 4** Structure of the vision model.

For classification using CNNs, to see the regions of interest along with the output description, we used Grad-CAM to obtain a saliency map [17]. When a CNN is used for classification into two classes, the background area, where no object exists, can sometimes have a high value on the saliency map. To focus on the target benign and malignant cells for analysis, we added one category for no cells, similar to object detection models [18]. In addition to benign and malignant images, 300 background images without lung cells were prepared, and a CNN was trained to classify these three categories. The benign and malignant cell probabilities were compared and the category with the highest probability was used as the classification result.

In the training of each CNN model used in this study, we used the Adam optimization algorithm, a learning coefficient of $1.0 \times 10^{-5}$, a batch size of 16, and a fixed number of 50 training epochs, with early stopping applied if the validation error did not improve. A CNN classifies images into benign, malignant, and background. The features obtained from the intermediate layer were used in the text decoder.

**D. Text Decoder**

The text decoder outputs the text based on the image features provided by the vision model. Before being input into the text decoder, the text provided as training data was divided into tokens by separating them with spaces and punctuation marks. Then, each token was converted into an integer value using a vectorizer, which also provided positional information to vectorize the text. Next, the text was decoded using a Transformer with multiple Transformer layers, as shown in Fig. 5. Each layer consisted of a multi-headed causal self-attention layer, a cross-attention layer, and a fully connected layer with 512-256 units. Finally, the output-vectorized text was converted into words using the reverse tokenization procedure to obtain the final text.

In this study, two text decoders were implemented, each specialized for either benign or malignant cases. The rationale for splitting the text decoders lies in the distinct nature of the textual descriptions required for these categories. Benign cases typically involve simpler descriptions, such as the types of cells and the condition of the background observed in the image. In contrast, malignant cases require more detailed descriptions, focusing on the specific characteristics of malignant cells. Based on these differences, we hypothesized that separating the text decoders would enable more stable and accurate text generation compared to using a single decoder for both types of cases. To achieve this, the vision model's classification results were used to switch between the two decoders. Each decoder was trained to generate descriptions only for benign and malignant cells, and the learning parameters for the text decoder were the Adam optimization algorithm and a learning coefficient of $1.0 \times 10^{-4}$. Regarding the number of training epochs, training was terminated if the validation error did not improve after 100 epochs.

Furthermore, the text produced by this method is case-insensitive and contains no punctuation. Therefore, we used the pretrained T5 model [19] to correct capitalization and punctuations in the sentences.

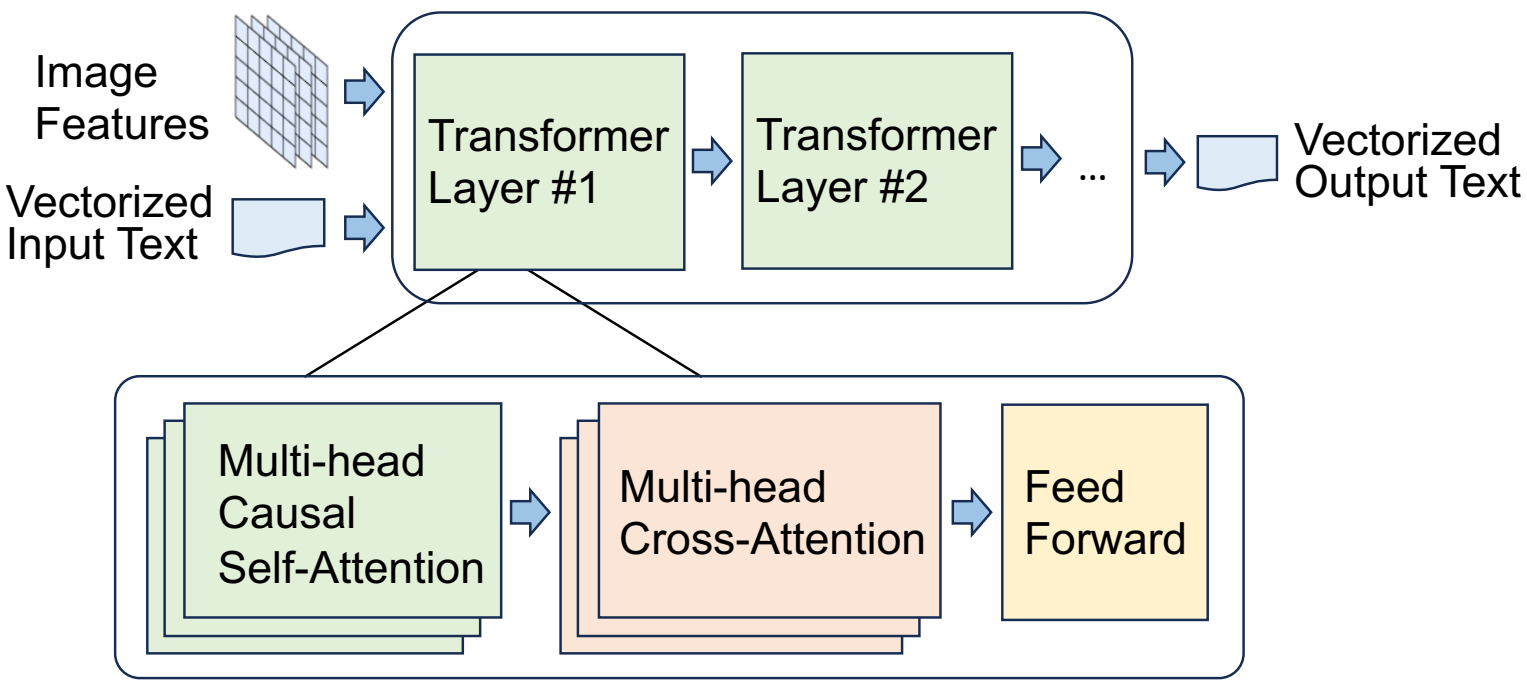

**Fig. 5** Text decoder structure. The Transformer layer consists of a multi-head causal self-attention layer, a cross-attention layer, and a feed forward network. The output is the vectorized text.

**E. Evaluation Metrics**

We evaluated the classification performance of the CNNs used in the proposed method and the quality of the description-generation model. First, to evaluate the CNN performance for classifying benign and malignant cases, we performed five iterations with multiple CNN-based SOTA models and calculated their detection sensitivity, specificity, and average (balanced accuracy). Based on the results, we selected the CNN-based model to be used in the vision model of the proposed method.

Next, we evaluated the image-captioning techniques by assessing the agreement between the ideal and generated descriptions using five metrics: bilingual evaluation understudy (BLEU) [20], metric for evaluation of translation with explicit ordering (METEOR) [21], recall-oriented understudy for gisting Evaluation (ROUGE) [22], consensus-based image caption evaluation (CIDEr) [23], and semantic propositional image caption evaluation (SPICE) [24]. These indices allow the evaluation of the degree of agreement between the ideal text and the generated text and take a range between 0 and 1. Larger values of these indicators indicate better agreement. For BLEU, n-grams up to 4 (BLEU-4) were used to calculate precision, comparing n-grams in the generated text with those in the gold standard. METEOR evaluated word matches while accounting for synonyms and alignment. ROUGE focused on recall-based overlaps of n-grams between the generated and ideal texts. CIDEr incorporated term frequency-inverse document frequency (TF-IDF) weighting to emphasize the importance of specific terms. SPICE assessed the semantic consistency of generated descriptions by comparing their scene graph representations to those of the gold standard.

The number of layers and heads of the Transformer used for text decoding can vary, and the performance of the proposed method changes accordingly. Therefore, we evaluated the performance of the proposed method by varying the number of layers between one, two, and three, and the number of heads between two, four, and six.

As an ablation study, we also built a single-text decoder model based on the proposed method, shared by both benign and malignant cells, and compared the performance of the proposed method's Transformer. The aforementioned number of heads and layers of the Transformer were optimized in the same manner as in the proposed method.

In addition, we fine-tuned the major captioning models, namely, GIT [25], BLIP [26], and BLIP2 [27], using large natural language models (LLMs) and compared them with the proposed method. For GIT and BLIP, we used the Base and Large models, while for BLIP2, we used a model incorporating OPT [28] with parameters of 2.7B and 6.7B as the LLM. For these processes, a PC with an NVIDIA RTX 6000Ada GPU and an AMD Ryzen9 CPU was used.

## III. RESULTS

Table 1 presents the classification performance evaluation results for the CNN used as the vision model. In this evaluation, fivefold cross-validation was conducted five times using the training data, and the average and standard deviations of the sensitivity, specificity, and balanced accuracy were obtained. By comparing the classification performance of the multiple CNN models, it was found that the balanced accuracy did not differ significantly among the models. Considering the importance of sensitivity in classifying benign and malignant cases, it was determined that ResNet50 was the best CNN model for use as a vision encoder.

Subsequently, ResNet50 was trained using all the training data prepared in this study, and the confusion matrix of the classification procedure was calculated using the test data, as presented in Table 2. The sensitivity and specificity for detecting malignant cells were 100% and 98.1%, respectively. Regarding to the background images, they were correctly classified as background.

Next, we processed and evaluated the entire description-generation method, using ResNet50 as the CNN for the vision model and changing the number of layers and heads used in the Transformer. Table 3 presents the evaluation accuracy results for the descriptions generated using BLEU, ROUGE, METEOR, CIDEr, and SPICE. As the proposed method uses different text decoders for benign and malignant cells, the evaluation was divided into benign and malignant cell cases. For benign cells, the smallest model with one layer and two heads exhibited the best generated text output, whereas for malignant cells, the model with two layers and four heads exhibited the best performance.
Table IV presents the evaluation results for the proposed method, the single-text decoder prepared as an ablation study, and the existing SOTA image-captioning models, including GIT, BLIP, and BLIP2. The Hugging-Face library was used to implement the captioning process using GIT, BLIP, and BLIP2.

Finally, Fig. 6 shows the proposed method output, including the input images, saliency maps using Grad-CAM, image classification results, and description output. For the output text, in addition to the proposed method, the results of the single-text decoder from the ablation study and the BLIP Base model, which exhibited the best performance among the existing captioning models, are also shown.

**Table 1** Classification performance of the CNNs for the vision model

| CNN model | Sensitivity | Specificity | Balanced accuracy |
|---|---|---|---|
| ResNet50 | 0.981±0.019 | 0.938±0.046 | 0.958±0.024 |
| VGG16 | 0.979±0.015 | 0.989±0.010 | 0.962±0.027 |
| InceptionV3 | 0.924±0.029 | 0.924±0.047 | 0.953±0.025 |
| DenseNet121 | 0.966±0.040 | 0.924±0.076 | 0.945±0.035 |

**Table 2** Classification confusion matrix (ResNet50)

| | | Predicted | |
|---|---|---|---|
| | | Benign | Malignancy |
| Actual | Benign | 54 | 1 |
| | Malignancy | 0 | 95 |

**Table 3** Network structure optimization in the text decoder.

| | | BLEU-1 | BLEU-2 | BLEU-3 | BLEU-4 | METEOR | ROUGE-L | CIDEr | SPICE |
|---|---|---|---|---|---|---|---|---|---|
| Layer 1 Head 2 | Benign | **0.959** | **0.950** | **0.941** | **0.932** | **0.683** | **0.966** | **8.509** | **0.946** |
| | Malignant | 0.850 | 0.798 | 0.749 | 0.714 | 0.487 | 0.841 | 4.598 | 0.715 |
| Layer 1 Head 4 | Benign | 0.961 | 0.950 | 0.939 | 0.928 | 0.671 | 0.965 | 8.170 | 0.942 |
| | Malignant | 0.849 | 0.796 | 0.746 | 0.712 | 0.485 | 0.835 | 4.419 | 0.684 |
| Layer 1 Head 6 | Benign | 0.950 | 0.938 | 0.926 | 0.914 | 0.653 | 0.950 | 7.976 | 0.920 |
| | Malignant | 0.852 | 0.797 | 0.744 | 0.709 | 0.486 | 0.837 | 4.651 | 0.723 |
| Layer 2 Head 2 | Benign | 0.959 | 0.949 | 0.939 | 0.928 | 0.675 | 0.966 | 8.340 | 0.943 |
| | Malignant | 0.847 | 0.791 | 0.736 | 0.699 | 0.481 | 0.835 | 4.459 | 0.717 |
| Layer 2 Head 4 | Benign | 0.944 | 0.926 | 0.908 | 0.888 | 0.623 | 0.943 | 7.103 | 0.906 |
| | Malignant | **0.853** | **0.804** | **0.756** | **0.724** | **0.494** | **0.844** | **4.883** | 0.718 |
| Layer 2 Head 6 | Benign | 0.950 | 0.933 | 0.917 | 0.901 | 0.637 | 0.946 | 7.539 | 0.925 |
| | Malignant | 0.853 | 0.800 | 0.749 | 0.714 | 0.489 | 0.842 | 4.735 | 0.710 |
| Layer 3 Head 2 | Benign | 0.954 | 0.939 | 0.923 | 0.905 | 0.639 | 0.956 | 7.297 | 0.922 |
| | Malignant | 0.824 | 0.767 | 0.710 | 0.672 | 0.462 | 0.814 | 4.025 | 0.674 |
| Layer 3 Head 4 | Benign | 0.949 | 0.938 | 0.927 | 0.916 | 0.668 | 0.957 | 8.230 | 0.938 |
| | Malignant | 0.852 | 0.799 | 0.748 | 0.713 | 0.488 | 0.842 | 4.704 | **0.721** |
| Layer 3 Head 6 | Benign | 0.940 | 0.924 | 0.908 | 0.891 | 0.627 | 0.941 | 7.273 | 0.902 |
| | Malignant | 0.841 | 0.787 | 0.736 | 0.700 | 0.480 | 0.833 | 4.563 | 0.702 |

In the Transformer for benign and malignant tumors, as shown in Fig. 5, the numbers of layers and heads were changed to assess the accuracy of the description. The indexes other than CIDEr have higher performance as they approach 1, and the larger the CIDEr, the higher the performance.

**Table 4** Performance comparison between the proposed method and SOTA models

| | | BLEU-1 | BLEU-2 | BLEU-3 | BLEU-4 | METEOR | ROUGE-L | CIDEr | SPICE |
|---|---|---|---|---|---|---|---|---|---|
| Proposed Method | Benign | 0.959 | 0.950 | 0.941 | 0.932 | 0.683 | 0.966 | 8.509 | 0.946 |
| | Malignant | **0.853** | **0.804** | **0.756** | **0.724** | **0.494** | **0.844** | **4.883** | 0.718 |
| | Balanced | **0.906** | **0.877** | **0.849** | **0.828** | **0.589** | **0.905** | **6.696** | 0.832 |
| CNN + single text decoder | Benign | 0.959 | 0.948 | 0.937 | 0.927 | 0.670 | 0.960 | 8.339 | 0.945 |
| | Malignant | 0.835 | 0.777 | 0.722 | 0.685 | 0.471 | 0.816 | 4.370 | 0.689 |
| | Balanced | 0.897 | 0.863 | 0.830 | 0.806 | 0.571 | 0.888 | 6.355 | 0.817 |
| GIT Base | Benign | 0.940 | 0.924 | 0.911 | 0.899 | 0.636 | 0.935 | 7.976 | 0.921 |
| | Malignant | 0.810 | 0.745 | 0.681 | 0.633 | 0.446 | 0.797 | 3.139 | 0.676 |
| | Balanced | 0.875 | 0.835 | 0.796 | 0.766 | 0.541 | 0.866 | 5.558 | 0.799 |
| GIT Large | Benign | 0.941 | 0.926 | 0.911 | 0.897 | 0.633 | 0.939 | 7.554 | 0.913 |
| | Malignant | 0.821 | 0.752 | 0.685 | 0.636 | 0.448 | 0.803 | 3.073 | 0.691 |
| | Balanced | 0.881 | 0.839 | 0.798 | 0.767 | 0.541 | 0.871 | 5.314 | 0.802 |
| BLIP Base | Benign | **0.971** | **0.962** | **0.952** | 0.942 | 0.688 | **0.975** | 8.339 | **0.952** |
| | Malignant | 0.835 | 0.777 | 0.719 | 0.672 | 0.467 | 0.828 | 3.010 | **0.717** |
| | Balanced | 0.903 | 0.870 | 0.836 | 0.807 | 0.578 | 0.902 | 5.675 | **0.835** |
| BLIP Large | Benign | 0.933 | 0.919 | 0.905 | 0.890 | 0.635 | 0.939 | 7.529 | 0.899 |
| | Malignant | 0.831 | 0.766 | 0.701 | 0.653 | 0.457 | 0.812 | 2.939 | 0.695 |
| | Balanced | 0.882 | 0.843 | 0.803 | 0.772 | 0.546 | 0.876 | 5.234 | 0.797 |
| BLIP2 2.7B | Benign | 0.405 | 0.393 | 0.380 | 0.366 | 0.517 | 0.642 | 0.856 | 0.909 |
| | Malignant | 0.844 | 0.783 | 0.721 | 0.671 | 0.469 | 0.833 | 2.879 | **0.739** |
| | Balanced | 0.625 | 0.588 | 0.551 | 0.519 | 0.493 | 0.738 | 1.868 | 0.824 |
| BLIP2 6.7B | Benign | 0.962 | 0.955 | 0.949 | **0.944** | **0.697** | 0.961 | **8.994** | 0.942 |
| | Malignant | 0.824 | 0.753 | 0.683 | 0.631 | 0.448 | 0.807 | 2.791 | 0.688 |
| | Balanced | 0.893 | 0.854 | 0.816 | 0.788 | 0.573 | 0.884 | 5.893 | 0.815 |

The results of the independent evaluation of benign and malignant cells and the average of both are labeled as Balanced.

**<Benign cells>**

Input image Saliency map

| Gold Standard | The background is relatively clean. There are small clusters of columnar epithelial cells. |
|---|---|
| Proposed | The background is relatively clean. There are small clusters of columnar epithelial cells. |
| Single decoder | The background is relatively clean. There are small clusters of columnar epithelial cells. |
| BLIP Base | the background is relatively clean there are small clusters of columnar epithelial cells |

Input image Saliency map

| Gold Standard | The background is relatively clean. There are small clusters of columnar epithelial cells. |
|---|---|
| Proposed | The background is relatively clean. There are small clusters of columnar epithelial cells. |
| Single decoder | There are adenocarcinoma cells with hyperchromatic nucleus, nucleus with irregular shape, pale cytoplasm nuclear enlargement. |
| BLIP Base | the background is relatively clean. there are small clusters of columnar epithelial cells. |

Input image Saliency map

| Gold Standard | The background is relatively clean. There is a squamous cell. |
|---|---|
| Proposed | The background is relatively clean. There are some squamous cells. |
| Single decoder | The background is relatively clean. There are some squamous cells. |
| BLIP Base | the background is relatively clean. there are a few columnar epithelial cells. |

Input image Saliency map

| Gold Standard | The background is relatively clean. There are clusters of columnar epithelial cells. |
|---|---|
| Proposed | There are adenocarcinoma cells with glandular structure or papillary cluster, hyperchromatic nucleus, nucleus with irregular shape, pale cytoplasm. |
| Single decoder | The background is relatively clean there are clusters of columnar epithelial cells |
| BLIP Base | the background is relatively clean. there are clusters of columnar epithelial cells. |

**<Malignant cells>**

Input image Saliency map

| Gold Standard | There are adenocarcinoma cells with glandular structure or papillary cluster, hyperchromatic nucleus, nucleus with irregular shape, pale cytoplasm. |
|---|---|
| Proposed | There are adenocarcinoma cells with hyperchromatic nucleus, nucleus with irregular shape, pale cytoplasm, nuclear enlargement, prominent nucleoli. |
| Single decoder | There are adenocarcinoma cells with hyperchromatic nucleus, nucleus with irregular shape, pale cytoplasm nuclear enlargement. |
| BLIP Base | there are adenocarcinoma cells with hyperchromatic nucleus, nucleus with irregular shape, pale cytoplasm, eccentric nucleus. |

Input image Saliency map

| Gold Standard | There are adenocarcinoma cells with hyperchromatic nucleus, nucleus with irregular shape, pale cytoplasm, nuclear enlargement, prominent nucleoli. |
|---|---|
| Proposed | There are adenocarcinoma cells with hyperchromatic nucleus, nucleus with irregular shape, pale cytoplasm nuclear enlargement. |
| Single decoder | There are adenocarcinoma cells with hyperchromatic nucleus, nucleus with irregular shape, pale cytoplasm nuclear enlargement. |
| BLIP Base | there are squamous cell carcinoma cells with thick cytoplasm, hyperchromatic nucleus, nucleus with irregular shape. |

Input image Saliency map

| Gold Standard | There are squamous cell calcinoma cells with thick cytoplasm, dark nucleus, nucleus with irregular shape. |
|---|---|
| Proposed | There are squamous cell carcinoma cells with thick cytoplasm, dark nucleus, nucleus, with irregular shape. |
| Single decoder | The background is relatively clean. There are clusters of columnar epithelial cells. |
| BLIP Base | there are squamous cell carcinoma cells with thick cytoplasm, hyperchromatic nucleus, nucleus with irregular shape. |

**Fig. 6** Images with saliency maps and corresponding generated descriptions with gold standard. The saliency map was obtained from ResNet50 in the proposed method. The underline in the text output indicates a major error.

## IV. DISCUSSION

In this study, we developed an image-classification and description-generation model for lung cytological images using feature extraction by a CNN and a text decoder using a Transformer. The fine-tuned CNN classifier exhibited acceptable performance with a classification accuracy of 95.8% for benign and malignant lung cells. ResNet50 was selected as the vision model based on its evaluation results, which showed the highest classification performance among the tested models. Its skip connections enable stable training even with increased depth, allowing it to effectively capture the complex morphological features of cytology images. Furthermore, its high computational efficiency supports accurate feature extraction, making it particularly suitable for this study. Additionally, the saliency map extracted by Grad-CAM highlighted typical benign and malignant cell areas effectively. These results confirm that feature extraction was properly performed. Based on the CNN classification results, two text decoders were used to generate descriptions on benign and malignant cells. In the proposed method, the initial CNN misclassified a benign cell as malignant in one image. This was the fourth benign image in Fig. 6, showing a clustered columnar epithelial cell. Although the cell appears slightly larger, it has cilia and shows no disarray in cell alignment, indicating it is a benign columnar cell. For this image, the proposed method generated the finding: "There are adenocarcinoma cells with glandular structure or papillary cluster, hyperchromatic nucleus, nucleus with irregular shape, pale cytoplasm." It is presumed that the model classified it as adenocarcinoma due to the nuclear shape and pale cytoplasm.

The text decoders consist of multiple Transformer layers with causal self-attention and cross-attention, and each attention layer has a multi-head attention structure. By optimizing the text decoder network structure by changing the number of layers and heads, it was found that the optimal decoder model that generated descriptions for benign cells was one with a single layer and two heads, while for the optimal text decoder for malignant cells had two layers and four heads. For benign cells, there were only two cell types, columnar and squamous epithelial cells; therefore, a simple network was sufficient. By contrast, malignant cells with a wide variety of cell types require more complex networks. The optimal network was selected based on the complexity of the target, yielding reasonable results.

Regarding the description output accuracy, the proposed method correctly provided a malignant description for all malignant cells and incorrectly provided malignant cells for only one case of benign cells. In contrast, the single-text decoder model evaluated in an ablation study and general captioning methods incorrectly output benign and malignant descriptions for some of the evaluated images. This difference in performance indicates that the proposed method has a CNN that classifies images into benign and malignant in the first stage, achieves high classification accuracy. The quantitative text evaluation metrics also demonstrate that the proposed method outperforms the SOTA LLM-based captioning models, indicating the effectiveness of the CNN for image classification and feature extraction. Additionally, regardless of the method used, benign cells achieved a higher agreement score than malignant cells.  This difference can be attributed to the nature of the descriptions required for each category. For benign cases, the generated text involves simpler descriptions, such as the types of cells and the condition of the background observed in the image. In contrast, malignant cases require more detailed descriptions to capture specific morphological features. Although the agreement scores for malignant cases were relatively lower than those for benign cases, they were still within an acceptable range. This demonstrates that the proposed method can handle the increased complexity of malignant descriptions while maintaining adequate performance. The difference in scores between benign and malignant cases reflects the inherent challenges posed by the variability and complexity of malignant cell morphologies.

In clinical cytology, benign and malignant cases are differentiated based on the morphology and arrangement of the cells, and these features are described as image findings; the differentiation results and image findings must be linked. Previous studies [10]-[12] on radiological and histological images used a single model, making it unclear how well the images were classified. This study combines a high-performance classification and text generation model, resulting in a reliable output. Experts reviewed the pairs of input images and the generated cytologic findings and confirmed that in many cases they matched. However, in some cases, such as the first example of a malignant cell in Fig. 6, there were slight discrepancies in the descriptions between the gold standard and the proposed method. Nonetheless, both descriptions referred to adenocarcinoma cells, and this difference was due to interpretational variation rather than a significant error. The image findings generated in this study have the potential to serve as reference information for diagnosis and report preparation in clinical settings.

This study has several limitations that should be addressed in future work. First, as a preliminary exploration into the generation of cytologic findings, it focused exclusively on typical benign and malignant cells. Expanding the scope to include all categories outlined in the WHO classification, such as atypical and suspicious for malignancy, will be critical for developing a more comprehensive and clinically applicable system. Achieving this will require more diverse datasets and robust modeling techniques to ensure accurate description generation across all categories. Incorporating larger datasets will not only improve model generalizability but also support the adoption of advanced vision models, such as Transformer-based architectures [29], which have shown promise in recent studies for extracting complex features from medical images and addressing the morphological variability present in cytological images. Another limitation of this study is its focus on lung cytology specimens. Applying this method to other cytological specimens, such as those from breast or thyroid tissues, may present unique challenges due to differences in cell morphology and staining characteristics. Future work should evaluate the adaptability of this methodology across various cytological domains to determine its generalizability and clinical utility. Finally, to address dataset limitations, it will be essential to digitize images using technologies such as z-stacks or extended focus with whole slide imaging (WSI) systems, rather than capturing individual images with a CCD camera as in this study. This approach will not only enable the collection of a larger and more diverse dataset but also help mitigate focus issues encountered during manual imaging.

## V. CONCLUSION

In this paper, we propose an image-classification and description-generation model for lung cytological images. The proposed method, which has a CNN that combines image classification and feature extraction and two Transformer-based text decoders, outperforms existing image capturing models and single-text-decoder models in terms of image identification and description-generation quality. These results indicate that the proposed method may be useful for generating cytological image findings.

**Author contributions:**

AT, TT, KI and HF conceived and designed the study. AM, YK and KI collected and assembled the data. AM, YK and TT prepared the text of cytologic finding. AT wrote the manuscript. HF and KI edited the manuscript. All authors, collectively accountable for all aspects of this work, participated in data analysis and gave final approval for the manuscript.

**Acknowledgements:**

Not applicable.

**Ethics approval statement:**

This study was conducted as a retrospective study with approval from the Institutional Review Boards of Fujita Health University (IRB No. HM23-390) and Meijo University (IRB No. 2023-44).

**Patient consent statement:**

Informed consent was obtained from all patients under data anonymization.

**Funding information:**

This work was supported in part by a Grant-in-Aid for Scientific Research (No. 23K07117), MEXT, Japan.

**Conflict of interest statement:**

All of the authors on this text state have no conflict of interest to declare.

**Data availability statement**

The data that support the findings of this study are available on request from the corresponding author. The data are not publicly available due to privacy or ethical restrictions.

## REFERENCES

[1] American Cancer Society, “Cancer facts and figures 2023”. Available at: https://www.cancer.org/content/dam/cancer-org/research/cancer-facts-and-statistics/annual-cancer-facts-and-figures/2023/2023-cancer-facts-and-figures.pdf.

[2] F. C. Schmitt et al., “The World Health Organization reporting system for lung cytopathology,” Acta Cytol., vol. 67, no. 1, pp. 80–91, 2023, doi: 10.1159/000527580.

[3] N. Thakur et al., "Recent Application of Artificial Intelligence in Non-Gynecological Cancer Cytopathology: A Systematic Review," Cancers, vol.14, 3529, 2022, doi: 10.3390/cancers14143529.

[4] A. Teramoto et al., “Automated classification of lung cancer types from cytological images using deep convolutional neural networks,” BioMed Res. Int., vol. 2017, pp. 4067832, 2017, doi: 10.1155/2017/4067832.

[5] T. Tsukamoto et al., “Comparison of fine-tuned deep convolutional neural networks for the automated classification of lung cancer cytology images with integration of additional classifiers,” Asian Pac. J. Cancer Prev., vol. 23, no. 4, pp. 1315–1324, 2022, doi: 10.31557/APJCP.2022.23.4.1315.

[6] A. Teramoto et al., “Automated classification of benign and malignant cells from lung cytological images using deep convolutional neural network,” Inform. Med. Unlocked, vol. 16, p. 100205, 2019, doi: 10.1016/j.imu.2019.100205.

[7] A. Teramoto et al., “Deep learning approach to classification of lung cytological images: Two-step training using actual and synthesized images by progressive growing of generative adversarial networks,” PLOS ONE, vol. 15, no. 3, p. e0229951, 2020, doi: 10.1371/journal.pone.0229951.

[8] D. Gonzalez et al., "Feasibility of a deep learning algorithm to distinguish large cell neuroendocrine from small cell lung carcinoma in cytology specimens," Cytopathology, vol. 31, no. 5, pp.426-431, 2020,

doi: 10.1111/cyt.12829.

[9] H. Park et al., "Deep Learning-Based Computational Cytopathologic Diagnosis of Metastatic Breast Carcinoma in Pleural Fluid," Cells, vol. 12, 1847, 2023, doi: 10.3390/cells12141847.

[10] X. Wang et al., “TieNet: Text-image embedding network for common thorax disease classification and reporting in chest X-rays” in, 2018 IEEE/CVF Conference on Computer Vision and Pattern Recognition. IEEE, Jun. 2018, 2018, pp. 9049–9058, doi: 10.1109/CVPR.2018.00943.

[11] B. Hou et al., “RATCHET: Medical Transformer for chest X-ray diagnosis and reporting” in Med. Image Comput. Comput. Assist. Interv. MICCAI, M. de Bruijne, P. C. Cattin, S. Cotin, N. Padoy, S. Speidel, Y. Zheng and C. Essert, Eds. Cham: Springer International Publishing, vol. 2021, pp. 293–303, 2021.

[12] Y. F. Zhou et al., “GNNFormer: A Graph-based Framework for Cytopathology Report Generation,” arXiv, Available at: arXiv:2303.09956, 2023.

[13] K. Simonyan and A. Zisserman, "Very deep convolutional networks for large-scale image recognition" in International Conference on Learning Representations, vol. 2015, 2015.

[14] C. Szegedy et al., “Going deeper with convolutions” in IEEE Conference on Computer Vision and Pattern Recognition (CVPR), vol. 2015, 2015, pp. 1–9, doi: 10.1109/CVPR.2015.7298594.

[15] K. He et al., “Deep residual learning for image recognition” in IEEE Conference on Computer Vision and Pattern Recognition (CVPR), vol. 2016, 2016, pp. 770–778, doi: 10.1109/CVPR.2016.90.

[16] G. Huang et al., “Densely connected convolutional networks” in IEEE Conference on Computer Vision and Pattern Recognition (CVPR), vol. 2017, 2017, pp. 2261–2269, doi: 10.1109/CVPR.2017.243.

[17] R. R. Selvaraju et al., "Grad-CAM: Visual explanations from deep networks via gradient-based localization" in IEEE International Conference on Computer Vision (ICCV), vol. 2017, 2017, pp. 618–626, doi: 10.1109/ICCV.2017.74.

[18] R. Girshick et al., "Rich feature hierarchies for accurate object detection and semantic segmentation" in IEEE Conference on Computer Vision and Pattern Recognition (CVPR), vol. 2015, 2014, pp. 580–587, doi: 10.1109/CVPR.2014.81.

[19] C. Raffe et al., "Exploring the Limits of Transfer Learning with a Unified Text-to-Text Transformer" The Journal of Machine Learning Research, vol. 21, no. 1, pp. 5485 - 5551, 2020.

[20] K. Papineni et al., "Bleu: A method for automatic evaluation of machine translation" in Annual Meeting of the Association for Computational Linguistics, pp. 1106–1114, 2001, doi: 10.3115/1073083.1073135.

[21] M. Denkowski and A. Lavie, "Meteor universal: Language specific translation evaluation for any target language" in EACL Workshop on Statistical Machine Translation, 2014, pp. 376–380, doi: 10.3115/v1/W14-3348.

[22] C. Y. Lin, "Rouge: A package for automatic evaluation of summaries" in Text Summarization Branches Out, 2004, pp. 74–81.

[23] R. Vedantam et al., "Cider: Consensus-based image description evaluation" in IEEE Conference on Computer Vision and Pattern Recognition, 2015, pp. 4566–4575, doi: 10.1109/CVPR.2015.7299087.

[24] P. Anderson et al., "Spice: Semantic propositional image caption evaluation" in European Conference on Computer Vision, 2016, pp. 382–398.

[25] J. Wang et al., "GIT: A Generative Image-to-text Transformer for Vision and Language," arXiv, Available at: arXiv:2205.14100v5, 2022.

[26] J. Li et al., "BLIP: Bootstrapping language-image pre-training for unified vision-language understanding and generation," Proc. 39th International Conference on Machine Learning, PMLR, vol. 162, pp. 12888-12900, 2022.

[27] J. Li et al., BLIP-2: Bootstrapping language-image pre-training with frozen image encoders and large language models," arXiv., Available at: arXiv:2301.12597v3, 2023.

[28] S. Zhang et al., "OPT: Open Pre-trained Transformer Language Models," arXiv, Available at: arXiv:2205.01068v4, 2022.

[29] A. Dosovitskiy et al., "An image is worth 16 × 16 words: Transformers for image recognition at scale," arXiv, Available at: arXiv:2010.11929, 2020.